\documentclass[10pt,twocolumn,letterpaper]{article}

\usepackage{cvpr}
\usepackage{times}
\usepackage{epsfig}
\usepackage{graphicx}
\usepackage{amsmath}
\usepackage{amssymb}
\usepackage{enumitem}

\PassOptionsToPackage{hyphens}{url}
\usepackage[colorinlistoftodos]{todonotes}

\usepackage{comment}
\usepackage{verbatim,epsfig,times,subfigure}
\usepackage{color}
\usepackage[export]{adjustbox}         
\usepackage[T1]{fontenc}

\DeclareMathOperator{\argmax}{argmax} 

\usepackage[breaklinks=true,bookmarks=false]{hyperref}

\cvprfinalcopy 

\def\cvprPaperID{****} 

\usepackage[ top = 1.50cm, bottom = 2.50cm, left = 2.00cm, right = 1.50cm]{geometry}

\begin{document}

\title{Deep Learning for Network Traffic Classification}

\author{Niloofar Bayat\\
Columbia University\\
{\tt\small niloofar.bayat@columbia.edu}
\and
Weston Jackson\\
Columbia University\\
{\tt\small wjj2106@columbia.edu}
\and
Derrick Liu\\
Columbia University\\
{\tt\small dl3122@columbia.edu}
}

\maketitle


\begin{abstract}
Monitoring network traffic to identify content, services, and applications is an active research topic in network traffic control systems. While modern firewalls provide the capability to decrypt packets, this is not appealing for privacy advocates. Hence, identifying any information from encrypted traffic is a challenging task. Nonetheless, previous work has identified machine learning methods that may enable application and service identification. The process involves high level feature extraction from network packet data then training a robust machine learning classifier for traffic identification. We propose a classification technique using an ensemble of deep learning architectures on packet, payload, and inter-arrival time sequences. To our knowledge, this is the first time such deep learning architectures have been applied to the Server Name Indication (SNI) classification problem. Our ensemble model beats the state of the art machine learning methods and our up-to-date model can be found on github: \url{https://github.com/niloofarbayat/NetworkClassification} 
\end{abstract}

\section{Introduction}\label{sec:intro}
Transport Layer Security (TLS) is one of the key cryptographic protocols for providing communication security over the Internet. The protocol allows client/server applications to communicate in a way that is designed to prevent eavesdropping, tampering, or message forgery \cite{rfc5246}. Today, TLS is central to the internet, and as it protects user privacy and security, websites are encouraged to use it. TLS is used extensively in HTTP, SMTP, FTP, and VoIP, where privacy and security is needed, and websites using HTTP in TLS tunnels (HTTPS) have increased drastically over the past decade \cite{naylor2014cost}. 

\begin{figure}
 \centering
  \includegraphics[width=0.45\textwidth]{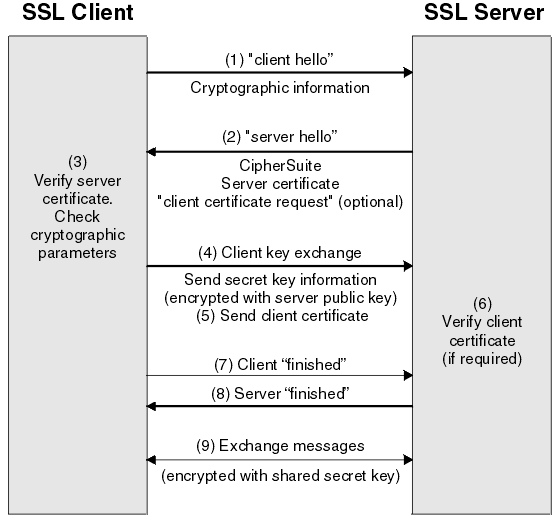}
 \caption{TLS handshake protocol}
 \label{fig:TLS}
\end{figure}

Over HTTPS services, the client and server first communicate through a TLS handshake, as shown in Figure \ref{fig:TLS}\footnote{This image is adopted from \url{https://www.ibm.com/support/knowledgecenter/en/SSFKSJ_7.1.0/com.ibm.mq.doc/sy10660_.htm}}. 
In this negotiation, protocol version, cryptographic algorithms, SSL certificates for authentication, and shared secrets based on public-key cryptography will be settled. If the handshake is completed successfully, client and server start to communicate information over an encrypted link \cite{rfc5246}.

Server Name Indication (SNI) is an extension to TLS handshake, which holds the destination hostname and can be extracted from Client-Hello message as in Figure \ref{fig:TLS}\cite{SNI}. SNI is a central component to HTTPS traffic inspection for many services and institutions. To preserve users' security, Firewalls inspect SNI to check if a server name is allowed. Moreover, intermediaries that censor their internet services also use SNI as a filter \cite{domain-fronting}. Since SNI is not encrypted, it does not completely preserve the privacy of users, and a man-in-the-middle can eavesdrop to discover the requested websites \cite{man-in-middle}. Moreover, SNI can be faked to bypass such Firewalls and eavesdroppers \cite{shbair2015efficiently}.
Since mid 2018, an upgrade called Encrypted SNI (ESNI) has been proposed to address this issue of domain eavesdropping \cite{domain-fronting, ESNI}. If successful, such a change would favor privacy advocates, leading to new challenges for network administrators and eavesdroppers alike. 

\section{Problem Formulation and Goals}

\begin{figure*}
 \centering
  \includegraphics[width=0.9\textwidth]{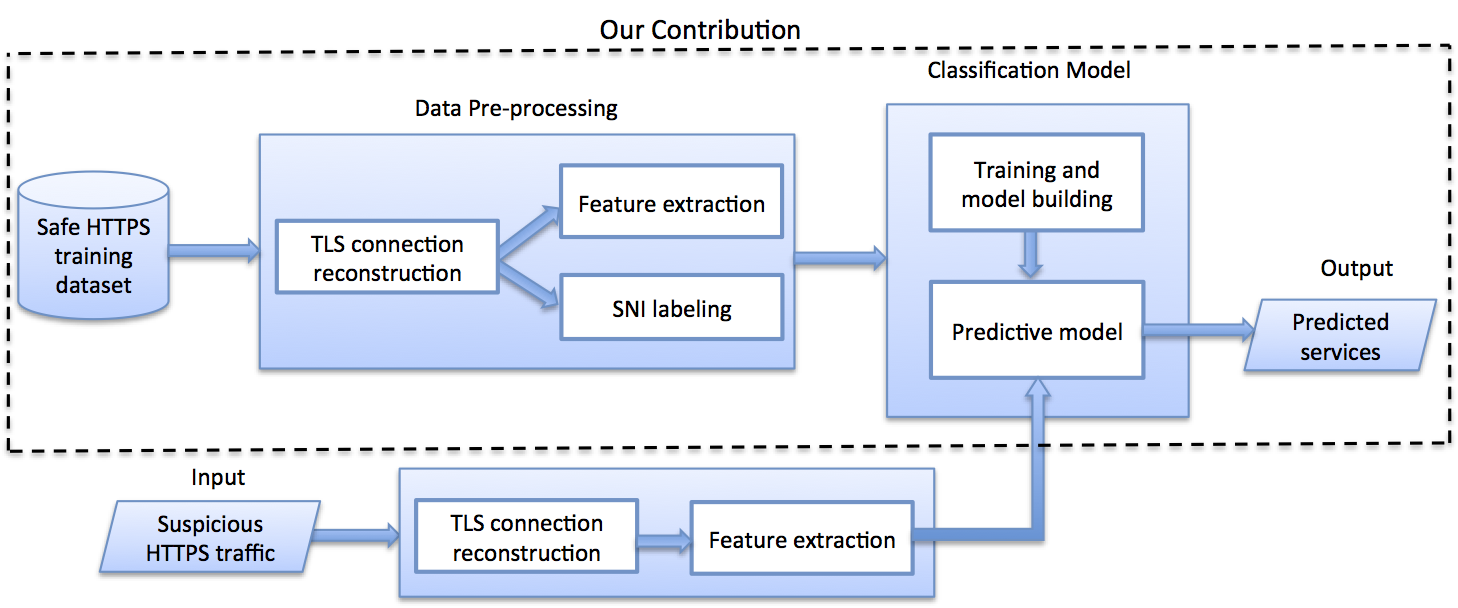}
 \caption{Workflow of the proposed HTTPS Identification model}
 \label{fig:Pipeline}
\end{figure*}

While ESNI would favor the privacy advocates over the eavesdroppers, previous work suggests that applications and services can be predicted with a high degree of accuracy without using SNI specific fields \cite{shbair2016multi, chen2010side}. Chen et al. find that side information can be extracted from many web applications despite using HTTPS protection \cite{chen2010side}. Their model assumes the eavesdropper can only observe the number of packets and timing/size of the packets. The information leaked includes health data, family income, and search queries. The root cause of these side-channel leaks is due to significant traffic and communication differences in web applications and defense from such leaks is non-trivial and application-specific \cite{chen2010side}. 

Traffic and communication differences between web applications could pose a significant threat to ESNI and other techniques to bypass SNI identification, as high SNI classification accuracy shows that such protocols are unable to completely protect user privacy from side-channel attacks. 

In this work, our main goal is to examine the effectiveness of deep learning for HTTPS SNI classification. We will only rely on encrypted TLS packet data without the SNI extension, and the SNI will constitute our ground truth labels. Under the assumption that SNI is not faked or forged, we will examine whether service identification accuracy can be improved with deep learning. To our knowledge, this is the first work to use deep learning on HTTPS data to classify SNI.

\section{Related Works}
\subsection{Machine Learning}
Previously, training supervised Naive Bayes classifiers as header-driven discriminators achieved high accuracy for many network traffic systems \cite{Moore:2005:ITC:1071690.1064220}. However, due to the growth of encrypted traffic, such approaches have been rendered ineffective. More recent approaches have focused on application level identification without using IP addresses, port numbers, or decrypted payload information. Alshammari et al. find that Decision Trees achieve the best accuracy when classifying Skype and SSH traffic \cite{5356534}. Okada et al. improve on this work by focusing on the creation of statistical features related to packet size and packet transfer times for application classification (FTP, DNS, HTTP, etc.) \cite{6147705}. These features achieve high accuracy when paired with Support Vector Machines classifiers. However, application level identification is not granular enough to address our research question, as our challenge is to detect the underlying service name rather than the type of traffic. 

Shabir et al. were one of the first to tackle this more granular problem of service identification for HTTPS-specific traffic (for instance maps.google.com vs drive.google.com) \cite{shbair2016multi}. Their work includes collecting HTTPS traces from user sessions and using the SNI extension for labelling each connection. Their proposed statistical framework includes the standard packet and inter-arrival time statistics, as well as additional statistical features related to the encrypted payload. They achieve their best results using Decision Tree and Random Forest classifiers. 

\subsection{Deep Learning}

There are currently few papers that apply deep learning to network traffic classification problems. Lopez-Martin et al. appear to be the first to apply Recurrent Neural Networks (RNN) and Convolutional Neural Networks (CNN) to the application level identification problem \cite{8026581}. Their CNN-LSTM architecture uses source port, destination port, packet size, TCP window size, and inter-arrival times as features, and beats the standard Random Forest classifier. Importantly, they also find that a large number packets is not necessary, as between 5-15 packets is sufficient to achieve excellent results. We find their work helpful in its analysis of deep learning architectures for network traffic, as many of their conclusions can also be applied to the SNI classification problem.

\section{Methods}

\begin{figure}
 \centering
  \includegraphics[width=0.45\textwidth]{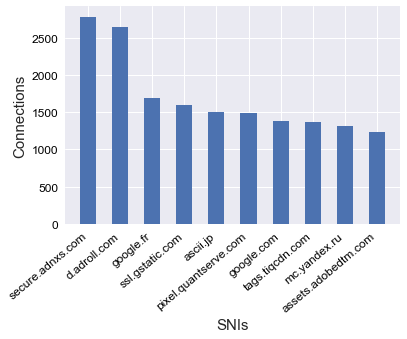}
 \caption{Top 10 SNIs with connection counts for day one Google Chrome data.}
 \label{fig:classes.png}
\end{figure}

\subsection{Data Collection and Labeling}

Publicly available HTTPs data was gathered by crawling top accessed HTTPs websites twice a day on both Google Chrome and Mozilla Firefox \cite{shbair2016}. This data was gathered over a two week period in 2016, and consists of almost 500,000 HTTPs flows from thousands of different services and websites. The dataset is made up of twenty-four raw packet capture (pcap) files, each between four and six gigabytes of data. We restrict our paper to Google Chrome data only, which consisted of 301,018 HTTPS flows (about 25GB of data). 

To generate our training data, we perform various preprocessings on the pcap files. Since a pcap file may contain all different kinds of traffic within a local machine, we use SSL filter on Wireshark to obtain only the HTTPS traffic \cite{shbair2016multi}. Then, since we are interested in packets in both directions (from the local machine to a specific server and vice versa), 
we wrote a script to distinguish incoming versus outgoing packets. Our method involves forming a 4-tuple for each TCP connection, consisting of source IP, destination IP, source port number, and destination port number. We unify the TCP connections which have their source and destination IP/port reversed, since that implies two directions of communication with a specific server. We then filter out all unknown SNIs and clean the remaining labels by removing numbers, dashes, and other unnecessary characters. 

For each connection, we store the following attributes in memory: SNI (label), accumulated bytes, arrival times, packet sizes, and payload sizes. Figure \ref{fig:classes.png} shows the most common SNI labels and their total number of connections. Once the pcap files are loaded into memory, we generate the datasets to be used for classification.

\subsection{Statistical Features}
The first dataset generated from the pcap files consists of forty-two statistical features used for training and validating the standard Random Forest classifier as presented by previous work \cite{shbair2016multi}. The features are as follows, and for each group of features we calculate \emph{remote}$ \rightarrow$\emph{local}, \emph{local}$\rightarrow$\emph{remote}, and \emph{combined} directions of communication\footnote{Payload size only uses \emph{remote}$ \rightarrow$\emph{local}, \emph{local}$\rightarrow$\emph{remote} as in \cite{shbair2016multi}}. 
\begin{itemize}[noitemsep]
\item Packet size: \{num, 25th, 50th, 75th, max, avg, var\}
\item Inter-arrival time: \{25th, 50th, 75th\}
\item Payload size: \{25th, 50th, 75th, max, avg, var\}
\end{itemize}

\subsection{Sequence Features}

Our second dataset consists of the sequences of packet sizes, payload sizes, and inter-arrival times generated from the TLS handshake that are needed for our Recurrent Neural Network. These features were chosen as they are standard features for machine learning network traffic classification, and each feature consists of a meaningful sequence. We use sequences from the combined packets for our training data, instead of sequences from \emph{local}$\rightarrow$\emph{remote} or \emph{remote}$\rightarrow$\emph{local}. Each length-$n$ sequence corresponds with the first $n$ packets per TCP connection, ordered by arrival time. All shorter sequences are pre-padded with zeros such that each input has the same length.

Importantly, because inter-arrival times have large variance even within the same sequence (anywhere from milliseconds to several minutes), directly training an RNN on inter-arrival time sequences yields poor results. While standardizing and normalizing leads to improvement in some cases, it does not solve the issue of a single inter-arrival time $t$ being several orders of magnitude larger than every other time unit in the sequence. Moreover, the largest inter-arrival times are often near the end of the packet sequence, which can adversely affect how the gradient propagates through time. We mitigate this problem by training our RNN on $\log{t}$. This decision greatly improves our results as using $\log{t}$ preserves similarity in the time sequence while not letting a single large inter-arrival time throw off the gradient calculation.

Choosing the correct packet sequence length is a balancing act. The longest TCP handshakes in our dataset have tens of thousands of packets. This makes training an RNN on full sequences extremely time-consuming. However, as in \cite{8026581}, we are able to get surprisingly good results with relatively small sequence lengths $n$. For all our results, we choose $n = 25$ as longer sequences yield diminishing accuracy improvements and much slower training times.

\subsection{Deep Learning}

Our deep learning architectures primarily make use of Convolutional Neural Networks (CNN) and Gated Recurrent Units (GRU) for sequence classification \cite{behnke2003hierarchical, ChungGCB14}. 

CNNs were initially applied to image processing and image classification, where feature engineering can be done automatically by extracting locational patterns from the image \cite{krizhevsky2012imagenet}. In our case, we apply a one dimensional CNN to our time series features to capture dependencies between feature vectors in consecutive time slots. 

GRUs are a type of Recurrent Neural Network that extend the conventional feedforward neural network to sequences of variable length \cite{ChungGCB14}. A GRU handles these sequences of variable length by holding a hidden state which has an activation dependent on the previous state. In this way, GRU is able to adaptively capture dependencies from time series data. 

\begin{figure}
 \centering \includegraphics[width=0.5\textwidth]{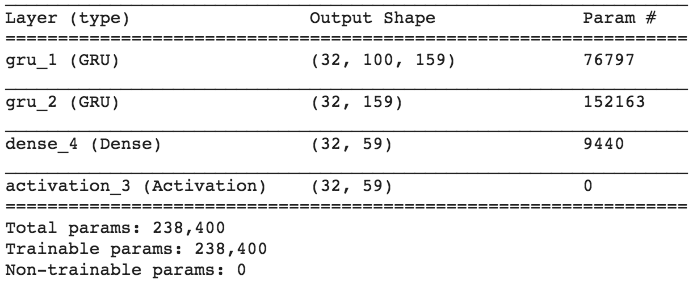}
 \caption{The summary of our preliminary model }
 \label{fig:model}
\end{figure}


\subsection{Evaluation Metrics}

We use several evaluation metrics for training and validating our classifiers. Our final reported results include accuracy, precision, recall, and F1-score for a variety of classifiers. For precision, recall, and F1-score, scikit-learn reports the macro-average which is an unweighted average of the calculated statistic per class. However, we calculate accuracy as simply as the sum of correct predictions over the total number of predictions. Thus, more prevalent classes have larger affect on our accuracy measurements. Ultimately, since the final metrics all return similar results, we will refer most often to accuracy.

All accuracy scores are reported using 10-Fold Cross Validation. Training and validation only includes SNI classes that meet a minimum number of connections (min connections) threshold. The rationale is that the training data needs a sufficient number of connections from a given SNI in order to learn the characteristics of the TLS handshake \cite{shbair2016multi}. As decreasing the min connections filter increases the number of classes, accuracy is typically lower with a lower min connections threshold. 

\subsection{Hardware}
Our preliminary results were run locally on MacOS with 16GB of memory. Our final training and validation results were run on a Google Deep Learning virtual machine. We use a Debian Linux OS instance with 16vCPUs and 60 GB of memory. 

\section{Results}
\subsection{Preliminary Model}
For our deep learning based architecture, we begin with an RNN trained on packet sequences only. The baseline architecture is a simple two layer GRU in Keras, which performs sequence to class modeling. We follow the two GRU layers with a fully-connected dense layer with Softmax activation. The output of the network has the number of neurons equal to the number of SNI classes. The neuron with the strongest activation represents the predicted class. The baseline architecture was trained for 10 epochs, and we use a batch size of 64, Adam optimizer, and sparse categorical cross entropy loss. A summary of the model can be seen in Figure \ref{fig:model}.

\begin{figure}[t]
\centering
{\includegraphics[width=0.45\textwidth]{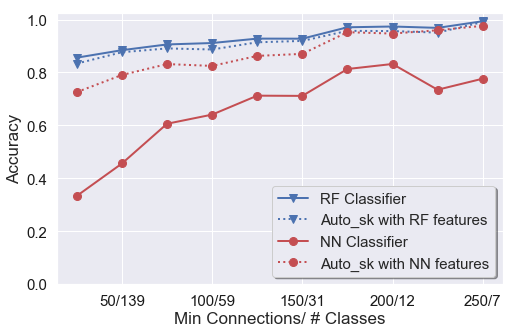}}
\caption{10-Fold Cross Validation accuracy on day one of data, while varying the minimum number of connections. Accuracy reported for Random Forest on statistical features, Auto-Sklearn on statistical features, baseline GRU RNN on sequence features, and Auto-Sklearn on sequence features. 
}
\label{fig:preliminaryResult}
\end{figure}

Figure \ref{fig:preliminaryResult} shows the accuracy for the Random Forest and Neural Net classifiers on the first day of TCP data. Note that the accuracy of the Random Forest classifier remains above 85\% even with a low barrier to entry (minimum connections $< 50$). Figure \ref{fig:preliminaryResult} also includes the accuracy of the proposed model from Auto-Sklearn \cite{NIPS2015_5872}. Auto-Sklearn is a library built off of sklearn that compares several supervised machine learning algorithms in order to automatically search for a optimal classifier with well tuned hyperparameters.

The Auto-Sklearn algorithm also selects a Random Forest classifier on the statistical features, which further supports the literature that claims Random Forest is ideal for network traffic classification on summary statistics. However, Auto-sklearn has a slightly lower accuracy than the original Random Forest classifier which is due to the lower number of estimators it chooses. This is simply due to the library's memory limits restrictions. Auto-Sklearn ensemble classifiers also outperform our baseline RNN classifier with packet size sequences as features. As the results suggest that the baseline RNN architecture has room for improvement, we iterate on this design in the following sections. 

\subsection{Improvements}
After our initial results, we make several important changes to improve the accuracy of our deep learning classifier. Our changes directly address two central problems pertaining to the baseline RNN: (1) The baseline RNN performs poorly on specific inputs, (2) The baseline RNN has high bias when there are many potential classes.

\begin{figure}
 \centering
  \includegraphics[width=0.5\textwidth]{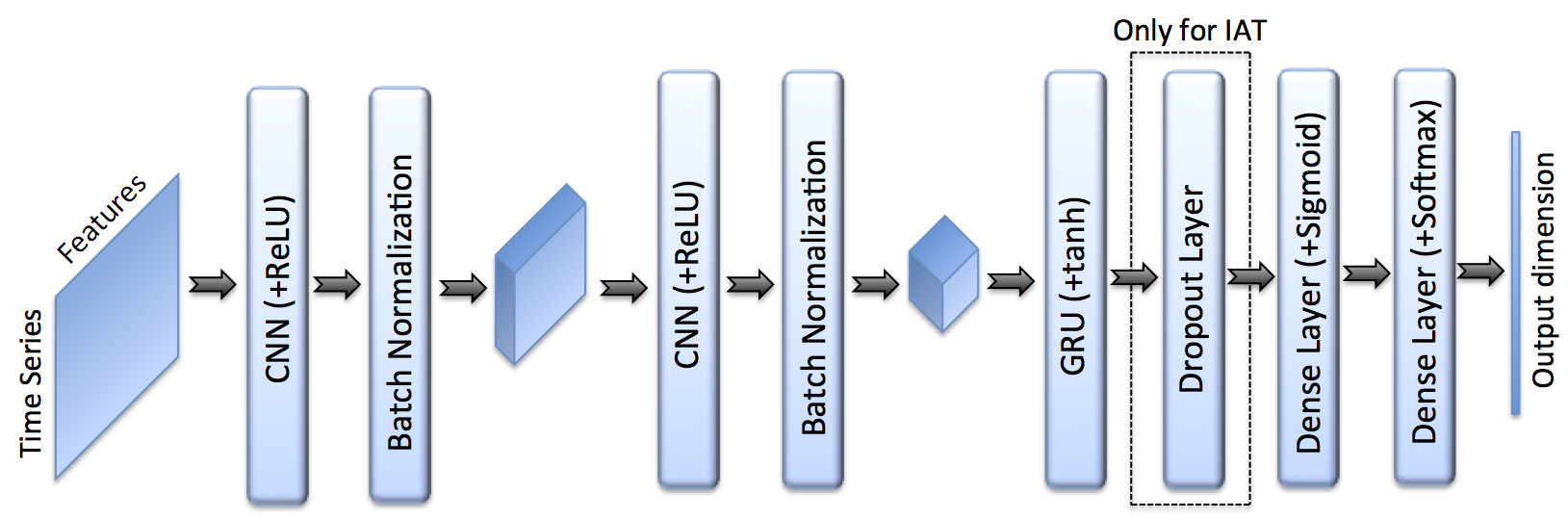}
 \caption{Layers of our final CNN-RNN model which uses inter-arrival time, payload length, and packet length represented in time-series as features. The first and second convolutional layers use size 3 kernel and 200 and 400 filters respectively. Our GRU uses 200 hidden units and our fully-connected layers have 200 and $n$ hidden units respectively, where $n$ is the number of classes. Dropout between the fully-connected layers is needed for inter-arrival time features to reduce over-fitting.}
 \label{fig:model_flowchart}
\end{figure}

To address the first problem, we employ the additional features (payload size, inter-arrival time) from the TCP-handshake to train the classifier. Yet rather than train a single classifier on all three features, we create separate classifiers to learn each feature for an ensemble method. Our rationale is that each deep learning architecture can be trained to recognize a different signal, such that an ensemble of all three is robust across many possible signals. Our final ensemble classifier simply chooses the class with the highest Softmax probability after averaging across the three individual classifiers. To address the high bias problem, we create a CNN-RNN architecture and add more complexity and hidden units to our RNN layers. An overview of our model is provided in Figure \ref{fig:model_flowchart}. We alter our model to include Convolutional layers, Batch Normalization, and an additional dense layer. To adjust for the much larger model, we add dropout to the inter-arrival time CNN-RNN which is prone to over-fitting. We also remove one GRU layer to speed up training time.

\begin{figure}[t]
\centering
{\includegraphics[width=0.48\textwidth]{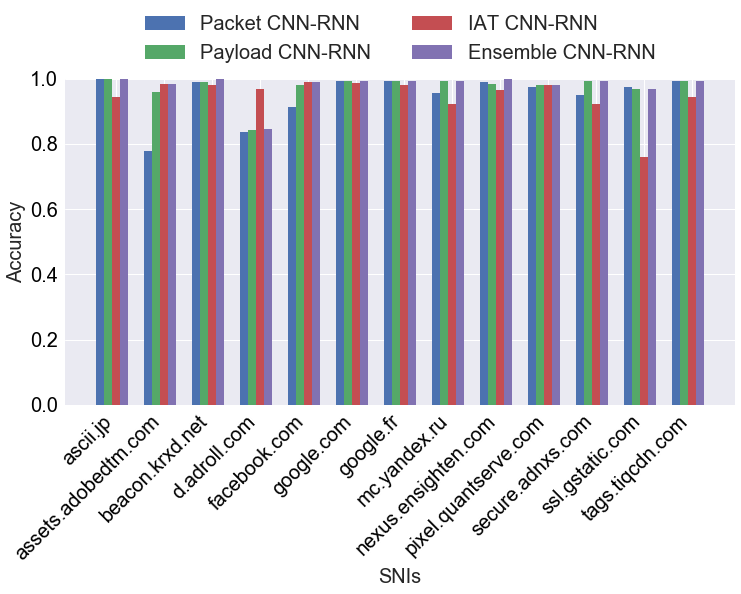}}
\caption{ A comparison of prediction accuracy for SNIs among different classifiers (1000 min connections). Ensembling the results of CNN-RNN classifiers, built from packet size, payload size and inter-arrival times leads to the best prediction accuracy.}
\label{fig:perSNIEnsemble}
\end{figure}

\begin{figure}[t]
\centering
\includegraphics[width=0.45\textwidth]{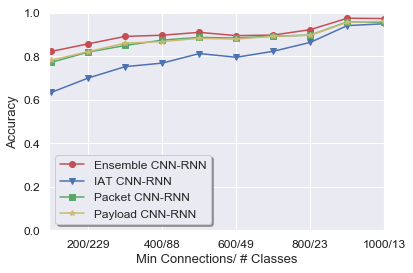}
\caption{Accuracy as a function of min connections for different classifiers: inter-arrival time trained CNN-RNN, packet trained CNN-RNN, payload trained CNN-RNN and ensemble CNN-RNN. As shown in the Figure, ensemble outperforms the rest. }
\label{fig:ensemble_classifier}
\end{figure}

The effectiveness of the ensemble method can be seen in Figure \ref{fig:perSNIEnsemble} which reports the accuracy of each classifier per-SNI with at least 1000 min connections, along with the accuracy of the ensemble classifier. While there are too many classes for a diagram at lower thresholds, Figure \ref{fig:ensemble_classifier} shows the performance boost from the ensemble method is clear across all minimum connections settings.

Finally, we make a few important changes to our training and validation process. Rather than train for 10 epochs, we use the early stopping feature provided by Keras to end training when validation loss does not improve for 5 epochs. Early stopping is a crucial component to our model's performance, as lower minimum connections thresholds can require 30+ epochs for validation accuracy to converge. 

\subsection{Best Results}
Our final results are reported after performing 10-Fold Cross Validation for a variety of minimum connections thresholds. The most important results are when the threshold is lowest (min connections = 100), as this is the most realistic and difficult scenario for a network traffic classifier. This is also the threshsold settings used for the same dataset in \cite{shbair2016multi}. Using 100 min connections leads to 532 possible SNI classes when we restrict ourselves to just Google Chrome data. 

For this setting, the Random Forest classifier achieves 92.2\% 10-Fold Cross Validation accuracy (see Appendix). Our preliminary baseline RNN trained on packet sequences achieves 67.8\% accuracy. The two-layer baseline CNN trained on packet sequences achieves 62.4 \% accuracy. The combined CNN-RNNs for packet, payload, and inter-arrival time sequences achieve 77.1\%, 78.1\%, and 63.2\% respectively (architectural changes alone lead to 10\% improvement for our deep learning classifier trained on packet sizes). Finally, the ensemble CNN-RNN achieves 82.3\% accuracy. Ultimately, with deep learning architectures alone, we are unable to improve upon the accuracy of the Random Forest classifier. 

Nonetheless, the Random Forest classifier can be beaten. To do so, we create an ensemble of the Random Forest with our best deep learning classifier. This requires us to average the Softmax output of our CNN-RNN ensemble with the log probability outputs of the Random Forest classifier provided by scikit-learn. Let $\sigma_{RF}(x)$, $\sigma_{d_1}(x)$, $\sigma_{d_2}(x)$, $\sigma_{d_3}(x)$ be the output probabilities for Random Forest, packet, payload, and inter-arrival time trained classifiers respectively. The final combined classifier amounts to choosing the SNI class with the highest average output across four total classifiers:\\

\begin{align*}
  \hat{y} &= \argmax \frac{1}{2}\sigma_{RF}(x)_y \\
  &+ \frac{1}{6}\sigma_{d_1}(x)_y \\
  &+ \frac{1}{6}\sigma_{d_2}(x)_y \\
  &+ \frac{1}{6}\sigma_{d_3}(x)_y
\end{align*}

\begin{figure}[t]
\centering
{\includegraphics[width=0.45\textwidth]{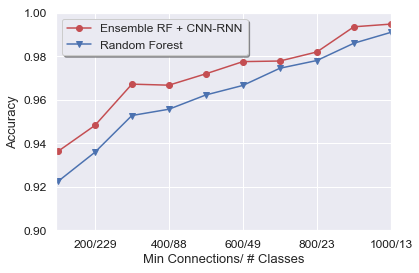}}
\caption{ Ensembling the results from inter-arrival time CNN-RNN, payload size CNN-RNN and packet size CNN-RNN, with Random Forest (note that the y-axis is restricted to [0.9, 1.0]). }
\label{fig:RFvsEnsemble}
\end{figure}

\begin{figure}[t]
\centering
{\includegraphics[width=0.48\textwidth]{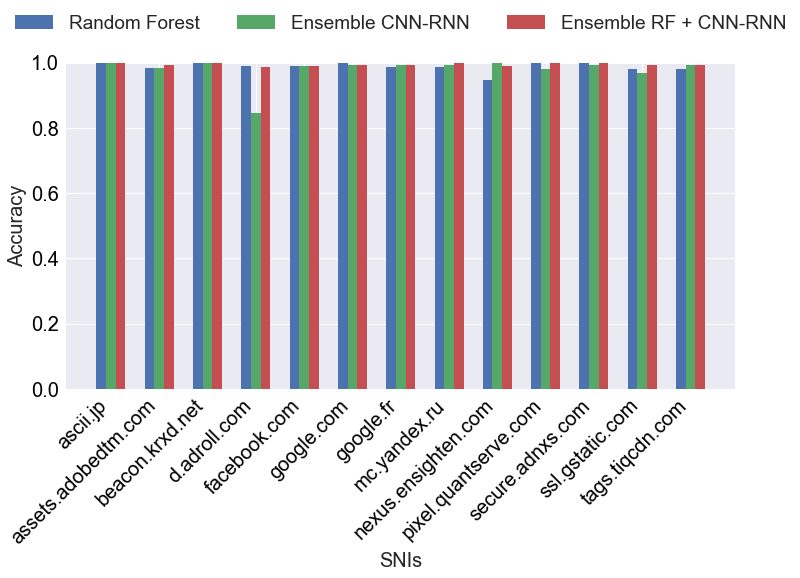}}
\caption{ A comparison of prediction accuracy for different SNIs as a function of different classifiers: Random Forest, ensemble CNN-RNN, and Random Forest + ensembled CNN-RNN. The plot clearly shows that ensembling leads to the best results. }
\label{fig:perSNIBest}
\end{figure}

\begin{figure}[t]
\centering
{\includegraphics[width=0.45\textwidth]{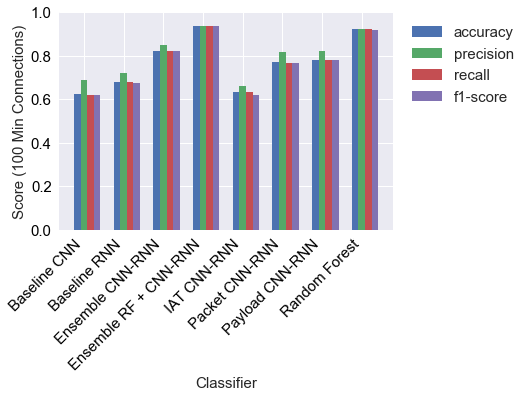}}
\caption{ Accuracy, precision, recall and f1-score for different classifiers (100 minimum connections). Ensemble of Random Forest with the newly built CNN-RNN network leads to improved accuracy/precision/recall and f1-score. } \label{fig:metrics_picture}
\end{figure}

Figure \ref{fig:RFvsEnsemble} shows how the ensemble Random Forest + CNN-RNN outperforms the Random Forest classifier at every tested minimum connection threshold. The rationale is similar to that of the previous section. While the Random Forest performs better on average, the CNN-RNN classifiers can outperform it on specific SNI-classes. For example, Figure \ref{fig:perSNIBest} shows that even with only 13 classes and a 1000 min connections threshold, ensembling the Random Forest with the CNN-RNN classifiers leads to better prediction on $\it{nexus.ensighten.com}$. A comparison of all classifiers for precision, recall, and F1-Score metrics can be seen in Figure \ref{fig:metrics_picture} as well as the Appendix. 

\section{Discussion}

\subsection{Architecture Variations}

Our deep learning architecture still has room for improvement, as we do not optimize the model for specific minimum connection thresholds. We use the same model for all settings, meaning that an architecture designed for a specific number of SNI-classes could outperform out current classifier. Additionally, the individual CNN-RNNs trained on packet, payload, and inter-arrival time sequences all use the roughly the same architecture. Further improvements are possible by optimizing each of these architectures individually, then combining the result into a stronger ensemble classifier. 

One possible area of improvement that we did test involves how the underlying CNN-RNNs are ensembled. For our best model, we weight the output of each CNN-RNN Softmax equally, but this is not necessarily ideal. Because packet and payload sizes are correlated, the CNN-RNNs trained on these features are learning a very similar signal. Thus, it is possible that an ensemble which gives more weight to the inter-arrival time classifier could outperform out current model. On the other hand, because the inter-arrival time CNN-RNN typically performs worse on lower minimum connections settings, it is also possible that decreasing its weight for the ensemble classifier could lead to better results. The appendix includes a few tests for ensembles with different combinations of features (packet + inter-arrival time, packet + payload, etc). Our results indicate that optimizing the combination of CNN-RNN outputs does not yield consistent improvement, and is largely dependent on the dataset and minimum connection thresholds on which the classifiers are trained. 

We leave optimizing a deep learning classifier for a specific number of SNI-classes and minimum connection settings as an open question and opportunity for future work.

\subsection{Directional Features}
Although \cite{8026581} suggests directional features can improve performance of a deep learning network traffic classifier, our results do not entirely reflect this. We conduct a test using an additional directional feature for our packet, payload, and inter-arrival time CNN-RNNs. We define the directionality $d$ of each packet in the packet sequence as follows: 
\[
d = \left\{
  \begin{array}{lr}
    1, &\text{client $\rightarrow$ server} \\
    -1, &\text{server $\rightarrow$ client } \\
    0, &\text{else (i.e. padding) }
  \end{array}
  \right\}
\]

\begin{figure}[t]
\centering
{\includegraphics[width=0.45\textwidth]{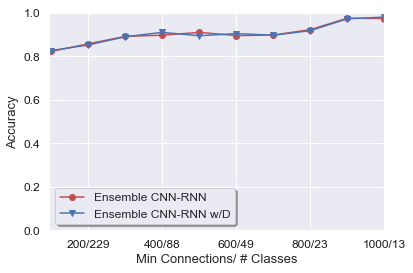}}
\caption{ Accuracy of ensemble CNN-RNN and ensemble CNN-RNN with additional directional features. As shown in the Figure, we do not find consistent improvement with the added features across minimum connections thresholds. Further results are in the Appendix. }
\label{fig:directionality}
\end{figure}

Figure \ref{fig:directionality} shows the accuracy of the final ensemble classifier with and without an additional direction vector added to each CNN-RNN. Our results indicate inconsistent improvement with this additional feature across minimum connection thresholds. Accuracy results for the individual CNN-RNNs with directionality are also reported in the Appendix. Ultimately, we believe that the use of directionality as a feature for deep learning network traffic classifiers needs to be further investigated. 

\section{Conclusion}
This work adds to the literature of TLS-based encrypted traffic classification by applying a deep neural network architecture for SNI detection. While most existing work addresses identifying application types, we focus on identifying HTTPS services. Our motivation for investigating SNI classification is twofold: (1) there are currently active research projects dedicated to eliminating the SNI extension to prevent eavesdropping \cite{domain-fronting, ESNI}, (2) even in existing systems, SNI can be spoofed by users \cite{shbair2015efficiently}. However, assuming SNI can be predicted from encrypted traffic with high accuracy, such techniques may be rendered ineffective.

Using a neural network architecture, we are able to effectively identify SNI by only considering statistics and sequences of encrypted TCP traffic, without decryption or using any header information. Our model consists of a combination of Recurrent Neural Networks, Convolutional Neural Networks, and Random Forest (the best machine learning method based on literature and Auto-Sklearn learn). By carefully analyzing different methods and studying the most informative features of network flow data, we achieve a high accuracy for an ensemble model which, to the best of our knowledge, outperforms the state of art. Future work would be using this model on real-time HTTPS traffic to test its effectiveness in predicting internet services. 

\section{Acknowledgements}
We would like to acknowledge Professor Drori for his advice and guidance as well as the authors W.M Shbair et al. for sharing their previous work.
\bibliographystyle{unsrt}
\bibliography{main}

\begin{thebibliography}{10}

\bibitem{rfc5246}
T.~Dierks.
\newblock The transport layer security (tls) protocol version 1.2 (rfc 5246).
\newblock \url{https://tools.ietf.org/html/rfc5246}, 2008.

\bibitem{naylor2014cost}
David Naylor, Alessandro Finamore, Ilias Leontiadis, Yan Grunenberger, Marco
  Mellia, Maurizio Munaf{\`o}, Konstantina Papagiannaki, and Peter Steenkiste.
\newblock The cost of the s in https.
\newblock In {\em Proceedings of the 10th ACM International on Conference on
  emerging Networking Experiments and Technologies}, pages 133--140. ACM, 2014.

\bibitem{SNI}
D.~Eastlake.
\newblock Transport layer security (tls) extensions: Extension definitions (rfc
  6066), 2011.

\bibitem{domain-fronting}
Thomas Claburn.
\newblock Don't panic about domain fronting, an sni fix is getting hacked out.
\newblock
  \url{https://www.theregister.co.uk/2018/07/17/encrypted_server_names/}, 2018.

\bibitem{man-in-middle}
Tanmay Patange.
\newblock Thow to defend yourself against mitm or man-in-the-middle attack.
\newblock
  \url{https://hackerspace.kinja.com/how-to-defend-yourself-against-mitm-or-man-in-the-middl-1461796382},
  2013.

\bibitem{shbair2015efficiently}
Wazen~M Shbair, Thibault Cholez, Antoine Goichot, and Isabelle Chrisment.
\newblock Efficiently bypassing sni-based https filtering.
\newblock In {\em Integrated Network Management (IM), 2015 IFIP/IEEE
  International Symposium on}, pages 990--995. IEEE, 2015.

\bibitem{ESNI}
SETH SCHOEN.
\newblock Esni: A privacy-protecting upgrade to https.
\newblock
  \url{https://www.eff.org/deeplinks/2018/09/esni-privacy-protecting-upgrade-https},
  2018.

\bibitem{shbair2016multi}
W.~M. Shbair, T.~Cholez, J.~Francois, and I.~Chrisment.
\newblock A multi-level framework to identify https services.
\newblock In {\em Network Operations and Management Symposium (NOMS), 2016
  IEEE/IFIP}, pages 240--248. IEEE, 2016.

\bibitem{chen2010side}
Shuo Chen, Rui Wang, XiaoFeng Wang, and Kehuan Zhang.
\newblock Side-channel leaks in web applications: A reality today, a challenge
  tomorrow.
\newblock In {\em 2010 IEEE Symposium on Security and Privacy}, pages 191--206.
  IEEE, 2010.

\bibitem{Moore:2005:ITC:1071690.1064220}
Andrew~W. Moore and Denis Zuev.
\newblock Internet traffic classification using bayesian analysis techniques.
\newblock {\em SIGMETRICS Perform. Eval. Rev.}, 33(1):50--60, June 2005.

\bibitem{5356534}
R.~Alshammari and A.~N. Zincir-Heywood.
\newblock Machine learning based encrypted traffic classification: Identifying
  ssh and skype.
\newblock In {\em 2009 IEEE Symposium on Computational Intelligence for
  Security and Defense Applications}, pages 1--8, July 2009.

\bibitem{6147705}
Y.~Okada, S.~Ata, N.~Nakamura, Y.~Nakahira, and I.~Oka.
\newblock Comparisons of machine learning algorithms for application
  identification of encrypted traffic.
\newblock In {\em 2011 10th International Conference on Machine Learning and
  Applications and Workshops}, volume~2, pages 358--361, Dec 2011.

\bibitem{8026581}
M.~Lopez-Martin, B.~Carro, A.~Sanchez-Esguevillas, and J.~Lloret.
\newblock Network traffic classifier with convolutional and recurrent neural
  networks for internet of things.
\newblock {\em IEEE Access}, 5:18042--18050, 2017.

\bibitem{shbair2016}
Wazen Shbair, Thibault Cholez, Jerome Francois, and Isabelle Chrisment.
\newblock Https websites dataset.
\newblock \url{http://betternet.lhs.loria.fr/datasets/https/}, 2016.

\bibitem{behnke2003hierarchical}
Sven Behnke.
\newblock {\em Hierarchical neural networks for image interpretation}, volume
  2766.
\newblock Springer, 2003.

\bibitem{ChungGCB14}
Junyoung Chung, {\c{C}}aglar G{\"{u}}l{\c{c}}ehre, KyungHyun Cho, and Yoshua
  Bengio.
\newblock Empirical evaluation of gated recurrent neural networks on sequence
  modeling.
\newblock {\em CoRR}, abs/1412.3555, 2014.

\bibitem{krizhevsky2012imagenet}
Alex Krizhevsky, Ilya Sutskever, and Geoffrey~E Hinton.
\newblock Imagenet classification with deep convolutional neural networks.
\newblock In {\em Advances in neural information processing systems}, pages
  1097--1105, 2012.

\bibitem{NIPS2015_5872}
Matthias Feurer, Aaron Klein, Katharina Eggensperger, Jost Springenberg, Manuel
  Blum, and Frank Hutter.
\newblock Efficient and robust automated machine learning.
\newblock In C.~Cortes, N.~D. Lawrence, D.~D. Lee, M.~Sugiyama, and R.~Garnett,
  editors, {\em Advances in Neural Information Processing Systems 28}, pages
  2962--2970. Curran Associates, Inc., 2015.

\end{thebibliography}

\begin{table*}[t]
\centering
{

\section*{Appendix}
\subsection*{Accuracy}
\begin{tabular}{ |c|c|c|c|c|c|c|c|c|c| } 
\hline
100 / 532 & 200 / 229 & 300 / 131 & 400 / 88 & 500 / 65 & 600 / 49 & 700 / 35 & 800 / 23 & 900 / 16 & 1000 / 13 \\ 
\hline
93.5\% A & 94.9\% A & 96.6\% A & 96.8\% A & 97.2\% A & 97.7\% A & 97.8\% A & 98.2\% A & 99.4\% A & 99.4\% A \\ 
92.2\% B & 93.7\% B & 95.2\% B & 95.6\% B & 96.2\% B & 96.8\% B & 97.5\% B & 97.7\% B & 98.7\% B & 99.1\% B \\ 
82.7\% D & 86.1\% F & 89.0\% D & 91.0\% D & 90.7\% C & 90.5\% D & 90.0\% E & 93.4\% F & 98.6\% F & 98.3\% G \\ 
82.2\% C & 86.0\% C & 89.0\% C & 90.1\% F & 89.5\% E & 89.6\% F & 89.9\% C & 93.3\% G & 98.4\% G & 98.1\% D \\ 
82.1\% G & 85.6\% G & 88.2\% G & 90.0\% C & 89.5\% D & 89.5\% C & 89.7\% D & 92.3\% C & 97.5\% C & 97.4\% C \\ 
81.0\% F & 85.2\% D & 88.1\% F & 89.7\% G & 89.4\% G & 89.3\% G & 89.6\% F & 91.8\% D & 97.3\% D & 97.3\% F \\ 
80.1\% E & 84.3\% E & 86.7\% E & 89.1\% H & 89.4\% F & 89.3\% I & 89.4\% I & 91.2\% E & 96.1\% J & 97.1\% I \\ 
78.5\% I & 82.5\% L & 85.9\% L & 89.0\% I & 88.8\% K & 88.6\% E & 89.3\% K & 90.3\% I & 96.0\% E & 96.9\% J \\ 
78.1\% L & 82.3\% K & 85.8\% I & 88.4\% E & 88.3\% L & 88.4\% K & 89.2\% L & 89.9\% L & 95.9\% K & 96.8\% E \\ 
77.1\% K & 81.6\% H & 85.2\% H & 87.8\% K & 88.3\% I & 88.1\% L & 89.0\% G & 89.9\% K & 95.8\% I & 96.7\% H \\ 
76.6\% H & 81.2\% I & 85.1\% K & 87.3\% L & 87.2\% N & 87.2\% H & 89.0\% H & 89.2\% H & 95.7\% L & 96.2\% L \\ 
67.8\% N & 79.3\% N & 81.8\% N & 85.2\% N & 87.1\% H & 86.0\% N & 86.7\% O & 87.8\% N & 95.6\% H & 95.8\% K \\ 
64.6\% J & 72.8\% J & 77.8\% O & 81.1\% O & 84.8\% O & 85.4\% O & 86.7\% N & 86.9\% J & 93.8\% N & 94.9\% N \\ 
63.2\% M & 72.4\% O & 74.9\% M & 79.5\% J & 81.2\% J & 80.0\% M & 83.4\% J & 86.5\% M & 93.7\% M & 94.9\% M \\ 
62.4\% O & 70.4\% M & 74.6\% J & 77.4\% M & 80.6\% M & 79.3\% J & 82.5\% M & 84.0\% O & 91.0\% O & 91.9\% O \\ 
\hline
\end{tabular}
}
\caption{ 10-Fold Cross Validation accuracies for all tested classifiers for given min connections / \# classes settings (specified in header). Results are from entire Google Chrome dataset. Precision, recall, and F1-Score are also reported on the next page. \newline\newline
Legend:\newline
A: Ensemble B + C \newline
B: Random Forest \newline
C: Ensemble K + L + M \newline
D: Ensemble H + I + J \newline
E: Ensemble K + L \newline
F: Ensemble K + M \newline
G: Ensemble L + M \newline
H: CNN-RNN (Packet + Directional features) \newline
I: CNN-RNN (Payload + Directional features) \newline
J: CNN-RNN (Inter-Arrival Time + Directional features) \newline
K: CNN-RNN (Packet) \newline
L: CNN-RNN (Payload) \newline
M: CNN-RNN (Inter-Arrival Time) \newline
N: Baseline RNN (Packet) \newline
O: Baseline CNN (Packet)
}
\end{table*}

\begin{table*}[t]
\centering
{
\subsection*{Precision}
\begin{tabular}{ |c|c|c|c|c|c|c|c|c|c| } 
\hline
100 / 532 & 200 / 229 & 300 / 131 & 400 / 88 & 500 / 65 & 600 / 49 & 700 / 35 & 800 / 23 & 900 / 16 & 1000 / 13 \\ 
\hline
93.8\% A & 95.0\% A & 96.6\% A & 97.0\% A & 97.0\% A & 98.0\% A & 98.0\% A & 98.0\% A & 99.0\% A & 99.4\% A \\ 
92.5\% B & 93.7\% B & 95.3\% B & 96.0\% B & 96.5\% B & 97.0\% B & 97.6\% B & 97.7\% B & 99.0\% B & 99.0\% B \\ 
86.0\% D & 88.2\% C & 90.6\% C & 92.0\% D & 92.0\% C & 92.0\% F & 93.0\% F & 94.0\% F & 99.0\% F & 98.0\% G \\ 
85.2\% C & 88.0\% F & 90.0\% F & 91.6\% C & 92.0\% D & 92.0\% D & 92.2\% C & 94.0\% G & 98.0\% G & 98.0\% D \\ 
84.0\% E & 88.0\% D & 90.0\% G & 91.0\% F & 91.0\% E & 91.4\% C & 92.0\% E & 93.7\% C & 98.0\% D & 97.8\% C \\ 
84.0\% F & 87.0\% E & 90.0\% D & 91.0\% G & 91.0\% F & 91.0\% G & 92.0\% G & 93.0\% E & 97.8\% C & 97.0\% E \\ 
84.0\% G & 87.0\% G & 89.0\% E & 91.0\% H & 91.0\% G & 91.0\% I & 92.0\% D & 93.0\% D & 97.0\% E & 97.0\% F \\ 
83.0\% I & 85.6\% L & 88.0\% L & 90.0\% E & 90.4\% K & 90.4\% K & 91.0\% H & 92.0\% I & 97.0\% H & 97.0\% J \\ 
82.2\% L & 85.2\% K & 88.0\% I & 90.0\% I & 90.0\% L & 90.0\% E & 91.0\% I & 91.8\% L & 97.0\% I & 97.0\% H \\ 
82.0\% H & 85.0\% H & 87.4\% K & 89.2\% K & 90.0\% I & 89.6\% L & 90.8\% K & 91.5\% K & 96.5\% K & 97.0\% I \\ 
81.6\% K & 85.0\% I & 87.0\% H & 88.8\% L & 89.0\% N & 89.0\% H & 90.8\% L & 91.0\% H & 96.2\% L & 96.6\% L \\ 
72.2\% N & 82.2\% N & 84.0\% N & 86.8\% N & 89.0\% H & 88.0\% N & 88.4\% N & 89.0\% N & 96.0\% J & 96.3\% K \\ 
69.0\% O & 78.0\% O & 81.0\% O & 83.0\% O & 86.0\% O & 87.0\% O & 88.0\% O & 89.0\% J & 94.6\% N & 95.3\% N \\ 
68.0\% J & 77.0\% J & 76.6\% M & 80.0\% J & 82.8\% M & 81.0\% M & 86.0\% M & 87.8\% M & 93.8\% M & 95.1\% M \\ 
66.2\% M & 73.4\% M & 76.0\% J & 78.8\% M & 82.0\% J & 79.0\% J & 85.0\% J & 85.0\% O & 93.0\% O & 93.0\% O \\ 
\hline
\end{tabular}

\subsection*{Recall}
\begin{tabular}{ |c|c|c|c|c|c|c|c|c|c| } 
\hline
100 / 532 & 200 / 229 & 300 / 131 & 400 / 88 & 500 / 65 & 600 / 49 & 700 / 35 & 800 / 23 & 900 / 16 & 1000 / 13 \\ 
\hline
93.6\% A & 95.0\% A & 96.4\% A & 97.0\% A & 97.0\% A & 98.0\% A & 98.0\% A & 98.0\% A & 99.0\% A & 99.4\% A \\ 
92.2\% B & 93.7\% B & 95.0\% B & 95.7\% B & 96.0\% B & 96.7\% B & 97.6\% B & 97.7\% B & 99.0\% B & 99.0\% B \\ 
83.0\% D & 86.0\% C & 89.0\% D & 91.0\% D & 90.6\% C & 90.0\% F & 90.0\% C & 93.0\% F & 99.0\% F & 98.0\% G \\ 
82.2\% C & 86.0\% F & 88.8\% C & 90.0\% C & 90.0\% E & 90.0\% D & 90.0\% E & 93.0\% G & 98.0\% G & 98.0\% D \\ 
82.0\% G & 86.0\% G & 88.0\% F & 90.0\% F & 89.0\% F & 89.6\% C & 90.0\% F & 92.3\% C & 97.5\% C & 97.3\% C \\ 
81.0\% F & 85.0\% D & 88.0\% G & 90.0\% G & 89.0\% G & 89.0\% E & 90.0\% D & 92.0\% D & 97.0\% D & 97.0\% E \\ 
80.0\% E & 84.0\% E & 87.0\% E & 89.0\% H & 89.0\% D & 89.0\% G & 89.5\% K & 91.0\% E & 96.0\% E & 97.0\% F \\ 
79.0\% I & 82.4\% L & 86.0\% I & 89.0\% I & 88.6\% K & 89.0\% I & 89.5\% L & 90.2\% L & 96.0\% J & 97.0\% J \\ 
78.2\% L & 82.2\% K & 85.8\% L & 88.0\% E & 88.2\% L & 88.6\% K & 89.0\% G & 90.0\% I & 96.0\% K & 97.0\% H \\ 
77.0\% H & 82.0\% H & 85.4\% K & 87.8\% K & 88.0\% I & 88.2\% L & 89.0\% H & 89.8\% K & 96.0\% H & 97.0\% I \\ 
76.8\% K & 81.0\% I & 85.0\% H & 87.2\% L & 87.2\% N & 87.0\% H & 89.0\% I & 89.0\% H & 96.0\% I & 96.2\% L \\ 
67.8\% N & 79.2\% N & 81.6\% N & 85.2\% N & 87.0\% H & 86.2\% N & 87.0\% O & 87.9\% N & 95.7\% L & 96.0\% K \\ 
65.0\% J & 73.0\% J & 78.0\% O & 81.0\% O & 85.0\% O & 85.0\% O & 86.6\% N & 87.0\% J & 93.8\% M & 95.0\% M \\ 
63.4\% M & 72.0\% O & 75.2\% M & 80.0\% J & 81.0\% J & 80.0\% M & 83.0\% J & 86.7\% M & 93.7\% N & 94.8\% N \\ 
62.0\% O & 70.4\% M & 75.0\% J & 77.8\% M & 81.0\% M & 79.0\% J & 82.2\% M & 84.0\% O & 91.0\% O & 92.0\% O \\ 
\hline
\end{tabular}

\subsection*{F1-Score}
\begin{tabular}{ |c|c|c|c|c|c|c|c|c|c| } 
\hline
100 / 532 & 200 / 229 & 300 / 131 & 400 / 88 & 500 / 65 & 600 / 49 & 700 / 35 & 800 / 23 & 900 / 16 & 1000 / 13 \\ 
\hline
93.6\% A & 95.0\% A & 96.4\% A & 97.0\% A & 97.0\% A & 98.0\% A & 98.0\% A & 98.0\% A & 99.0\% A & 99.4\% A \\ 
92.0\% B & 93.7\% B & 95.0\% B & 95.7\% B & 96.0\% B & 96.7\% B & 97.6\% B & 97.7\% B & 99.0\% B & 99.0\% B \\ 
83.0\% D & 86.0\% C & 89.0\% D & 91.0\% D & 90.6\% C & 90.0\% D & 90.0\% C & 93.0\% F & 99.0\% F & 98.0\% G \\ 
82.0\% C & 86.0\% F & 88.8\% C & 90.0\% C & 90.0\% E & 89.0\% C & 90.0\% E & 93.0\% G & 98.0\% G & 98.0\% D \\ 
82.0\% G & 85.0\% G & 88.0\% F & 90.0\% F & 89.0\% F & 89.0\% F & 90.0\% D & 92.0\% C & 97.5\% C & 97.3\% C \\ 
81.0\% F & 85.0\% D & 88.0\% G & 90.0\% G & 89.0\% G & 89.0\% G & 89.5\% K & 92.0\% D & 97.0\% D & 97.0\% E \\ 
80.0\% E & 84.0\% E & 87.0\% E & 89.0\% H & 89.0\% D & 89.0\% I & 89.5\% L & 91.0\% E & 96.0\% E & 97.0\% F \\ 
78.2\% L & 82.8\% L & 86.0\% I & 89.0\% I & 88.6\% K & 88.0\% L & 89.0\% F & 90.0\% K & 96.0\% J & 97.0\% J \\ 
78.0\% I & 82.4\% K & 85.8\% L & 88.0\% E & 88.2\% L & 88.0\% E & 89.0\% H & 90.0\% I & 96.0\% K & 97.0\% H \\ 
77.0\% H & 82.0\% H & 85.0\% H & 87.8\% K & 88.0\% I & 87.6\% K & 89.0\% I & 89.8\% L & 96.0\% H & 97.0\% I \\ 
76.8\% K & 81.0\% I & 84.8\% K & 87.2\% L & 87.0\% N & 87.0\% H & 88.0\% G & 89.0\% H & 96.0\% I & 96.2\% L \\ 
67.5\% N & 79.2\% N & 81.2\% N & 85.0\% N & 87.0\% H & 85.7\% N & 87.0\% O & 87.4\% N & 95.7\% L & 96.0\% K \\ 
63.0\% J & 73.0\% O & 78.0\% O & 81.0\% O & 85.0\% O & 85.0\% O & 86.6\% N & 87.0\% J & 93.8\% M & 95.0\% M \\ 
62.0\% O & 72.0\% J & 73.8\% M & 79.0\% J & 80.0\% J & 78.8\% M & 82.0\% J & 86.3\% M & 93.7\% N & 94.8\% N \\ 
62.0\% M & 69.0\% M & 73.0\% J & 76.8\% M & 79.6\% M & 78.0\% J & 81.2\% M & 84.0\% O & 91.0\% O & 92.0\% O \\ 
\hline
\end{tabular}}
\end{table*}

\end{document}